Advanced Structural
and Chemical Imaging

**RESEARCH**                                                      **Open Access**

# Optimal principal component analysis of STEM XEDS spectrum images

Pavel Potapov[1,2*] and Axel Lubk[2]

**Abstract**

STEM XEDS spectrum images can be drastically denoised by application of the principal component analysis (PCA). This paper looks inside the PCA workflow step by step on an example of a complex semiconductor structure consisting of a number of different phases. Typical problems distorting the principal components decomposition are highlighted and solutions for the successful PCA are described. Particular attention is paid to the optimal truncation of principal components in the course of reconstructing denoised data. A novel accurate and robust method, which overperforms the existing truncation methods is suggested for the first time and described in details.

## Background

Scanning transmission electron microscopy (STEM) delivers images of nanostructures at high spatial resolution matching that of broad beam transmission electron microscopy (TEM). Additionally, modern STEM instruments are typically equipped with electron energy-loss spectrometers (EELS) and/or X-rays energy-dispersive spectroscopy (XEDS, sometimes abbreviated as EDS or EDX) detectors, which allows to turn images into *spectrum-images*, i.e. pixelated images, where each pixel represents an EEL or XED spectrum. In particular, the recent progress in STEM instrumentation and large collection-angle silicon drift detectors (SDD) [1, 2] made possible a fast acquisition of large STEM XEDS spectrum-images consisting of 10–1000 million data points. These huge datasets show typically some correlations in the data distribution, which might be retrieved by application of statistical analysis and then utilized for improving data quality.

The simplest and probably the most popular multivariate statistical technique is a *principal component analysis* (PCA) that expresses available data in terms of orthogonal linearly uncorrelated variables called *principal* components [3–9]. In general terms, PCA reduces the dimensionality of a large dataset by projecting it into an orthogonal basic of lower dimension. It can be shown that among all possible linear projections, PCA ensures the smallest Euclidean difference between the initial and projected datasets or, in other words, provides the minimal least squares errors when approximating data with a smaller number of variables [10]. Due to that, PCA has found a lot of applications in imaging science for data compression, denoising and pattern recognition (see for example [11–18]) including applications to STEM XEDS spectrum-imaging [19–24].

A starting point for the PCA treatment is the conversion of a dataset into a matrix **D**, where spectra are placed on the matrix rows and each row represents an individual STEM probe position (pixel). Assume for definiteness that the $m \times n$ matrix **D** consists of $m$ pixels and $n$ energy channels. Although STEM pixels may be originally arranged in 1D (linescan), 2D (datacube) or in a configuration with higher dimensions, they can be always recasted into the 1D train as the neighborhood among pixels does not play any role in the PCA treatment. PCA is based on the assumption that there are certain correlations among spectra constituting the data matrix **D**. These correlations appear because the data variations are governed by a limited number of the *latent factors*, for example by the presence of chemical phases with the fixed composition. The spectral signatures of latent factors might be, however, not apparent as they are masked by noise. In this consideration, PCA relates closely to the *factor analysis* [7] although principal components generally do not coincide with the latent factors but represent rather their linear combinations [25].

*Correspondence: pavel@temdm.com
[1] Department of Physics, Technical University of Dresden, Dresden, Germany
Full list of author information is available at the end of the article

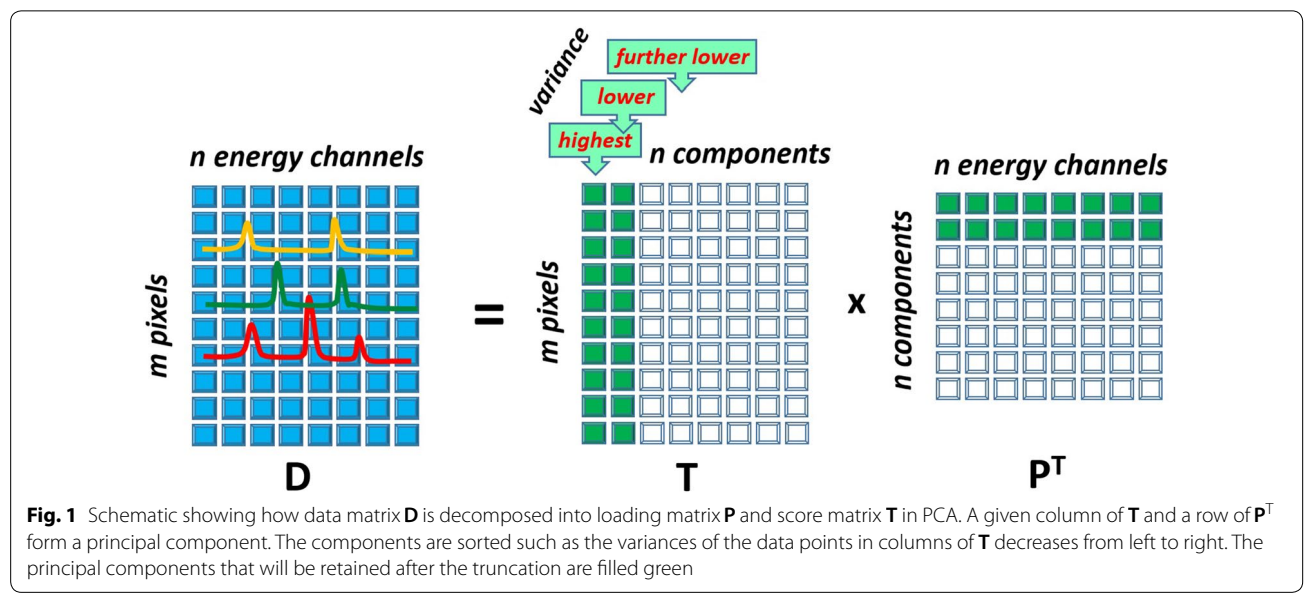

**Fig. 1** Schematic showing how data matrix **D** is decomposed into loading matrix **P** and score matrix **T** in PCA. A given column of **T** and a row of **P**$^T$ form a principal component. The components are sorted such as the variances of the data points in columns of **T** decreases from left to right. The principal components that will be retained after the truncation are filled green

Principal components can be found, for example, through the diagonalization of the covariance matrix **DD**$^T$ or by applying the singular value decomposition (SVD) directly to **D**. SVD decomposes data as:

$$\mathbf{D} = \mathbf{U}\mathbf{\Sigma}\mathbf{V}^T \quad (1)$$

where **U** and **V** are left and right hand singular vector matrices and **Σ** is a diagonal matrix with singular values of **D** on the diagonal. For the purpose of PCA formula (1) can be rewritten as:

$$\mathbf{D} = \mathbf{T}\mathbf{P}^T \quad (2)$$

where $\mathbf{P} = \mathbf{V}$ is an $n \times n$ *loading* matrix describing principal components and $\mathbf{T} = \mathbf{U}\mathbf{\Sigma}$ is an $m \times n$ *score* matrix showing the contribution of components into the dataset. Figure 1 illustrates the principal component decomposition in the graphical form. The columns of the loading matrix **P** (rows in **P**$^T$) represent spectra of principal components expressed in the original energy channels. Typically, the columns of **P** are normalized to unity, such that all the scaling information is moved into the score matrix **T**. It is important to sort the principal components in the order of their significance. In PCA, the components are ranked according their variance, i.e. the variance of the data along the corresponding column of **T**.

The data matrix **D** expressed by (1) can be subjected to dimensionality reduction or, in other words, truncation of components. Such dimensionality reduction might serve various purposes, for instance, it can be a first step for more complicated multivariate statistical treatment like unmixing data and extraction of latent factors. In the simplest case, dimensionality reduction can be utilized for removing the major part of noise from data, i.e. for its *denoising*.

The following questions are at the heart of the method. How much the dimensionality of a given dataset can be reduced? How many components must be retained to reproduce adequately the data variation and how many of them may be truncated to reduce noise? This paper attempts to address these crucial questions on the example of typical XEDS spectrum-images obtained in modern STEM instruments.

At a first glance, the reasonable number of retained components should be equal to the known (or expected) number of latent factors behind the data variations. It will be, however, shown that the situation is more complicated and the number of meaningful components might strongly differ from the number of latent factors—typically, there are less components than factors. The reason for this deviation is unavoidable corruption of data with noise.

To explore the topic most comprehensively, we considered an object with a very large number of latent factors and analyzed its experimental XEDS spectrum image. In parallel, we generated a twin synthetic object that mimicked the real one in all its essential features. An advantage of the synthetic data is the possibility to

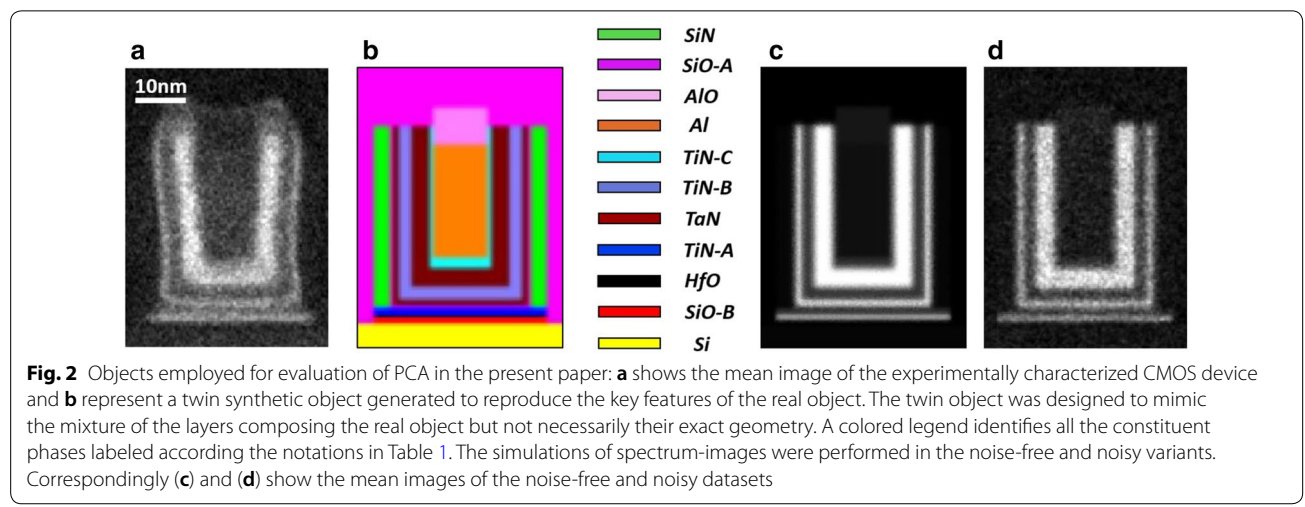

**Fig. 2** Objects employed for evaluation of PCA in the present paper: **a** shows the mean image of the experimentally characterized CMOS device and **b** represent a twin synthetic object generated to reproduce the key features of the real object. The twin object was designed to mimic the mixture of the layers composing the real object but not necessarily their exact geometry. A colored legend identifies all the constituent phases labeled according the notations in Table 1. The simulations of spectrum-images were performed in the noise-free and noisy variants. Correspondingly (**c**) and (**d**) show the mean images of the noise-free and noisy datasets

exclude noise in simulations and, therefore, compare the noisy data with the noise-free reference.

PCA is often considered as a fixed procedure, where little can be altered or tuned. In reality, there is a number of hidden issues hampering the treatment and leading to dubious results. Better understanding of the potential issues might help to design the optimal treatment flow improving the efficiency and avoiding artifacts. The systematic comparison between the experimental and synthetic data sets on the one hand and between the synthetic noisy set and the noise-free reference on the other hand, allowed us to identify the typical obstacles in the treatment flow and find the solutions for the optimal principal component decomposition and reconstruction of the denoised data.

Below, it will be demonstrated that certain pre-treatments, namely weighting datasets and reducing its sparseness, are essential for the successful PCA of STEM XEDS data. Ignoring these pre-treatments would deteriorate dramatically the denoising effect of PCA and might cause severe artefacts. This paper addresses also the problem of the optimal truncation of principal components in the course of reconstructing denoised data. A new accurate and robust method, which overperforms the existing truncation methods, is suggested and tested with a number of experimental and synthetic objects.

The paper is organized as follows: "Multi-component object for spectrum-imaging" section describes an object investigated with STEM XEDS and also its synthetic twin object designed to mimic the real one. "Principal component decomposition" section follows all steps of the principal component decomposition and highlights the potential problems distorting PCA in the case of XEDS spectrum images. "Truncation of principal components and reconstruction" section presents the theoretical background for truncation of principal components and discuss the existing practical truncation methods. A novel method for automatic determination of the optimal number of components is introduced in "Anisotropy method for truncation of principal components" section. At the end of "Truncation of principal components and reconstruction" section, the results of the spectrum-image reconstruction are shown and the denoising ability of PCA is demonstrated.

## Results and discussion

### Multi-component object for spectrum-imaging

A modern CMOS device was chosen as an object for XEDS spectrum imaging. Figure 2a shows a so-called mean image of the device, i.e. the spectrum imaging signal integrated over all available energy channels. Such images appear typically very similar to high-angle annular dark field (HAADF) images. The device consists of a number of nano-scale layers manufactured to optimize the speed, switching potential and leaking current of a field-effect transistor [26]. The chemical content of layers shows the high variety that makes such a device a good model object for the extensive PCA. The composition of the layers was measured as listed in Table 1 by Auger spectroscopy, ToF-SIMS and other non-TEM methods. There were 11 different layers, or in other words phases, although some phases differed only marginally in composition.

In parallel to experiment, we generated a twin synthetic object with the layers of the same composition and roughly the same relative volume fractions (Fig. 2b). As demonstrated below, the synthetic object shows a good proximity to the experimental one, which helps to figure out important regularities in its PCA treatment. Then XEDS spectrum-images of the synthetic object were generated in two variants: with and without adding a Poisson noise. These two datasets will be referred to as *noisy* and *noise-free* synthetic datasets in the present paper. The generated noisy and noise-free spectrum-images are presented in the DigitalMicrograph format in Additional files 1 and 2 respectively. The details of the experiment and simulations are described in "Experimental details" and "Details of simulation" subsections.

**Table 1 Composition of layers (phases) constituting the investigated CMOS device**

| Phase notation | Composition (at.%) |
| --- | --- |
| Si | 100% Si |
| SiO-A | 33% Si–67% O |
| SiO-B | 29% Si–57% O–14% N |
| HfO | 33% Hf–67% O |
| TiN-A | 50% Ti–50% N |
| TiN-B | 50% Ti–40% N–10% O |
| TiN-C | 45% Ti–45% N–10% Al |
| TaN | 50% Ta–50% N |
| Al | 80% Al–20% Ti |
| AlO | 40% Al–60% O |
| SiN | 43% Si–57% N |

## Principal component decomposition
### Unweighted PCA
In classical PCA, data is decomposed into principal components according to Eq. (2). Unfortunately, the corruption of data with noise makes such decomposition not perfect. The extracted eigenspectra of components always deviate from those for "true", noise-free data and the component variances are always overestimated due to the contribution of the noise variance. It is instructive to evaluate such deviations by comparison of the obtained noisy and "true" synthetic eigenspectra. In this paper, we introduce a proximity function $\phi$:

$$\phi_k(l) = \sum_{i=0}^{n} \left( p_{il} p_{ik}^* \right)^2 \tag{3}$$

that calculates a squared projection of $m$th column of noisy matrix $\mathbf{P}$ on the target $k$th column of noise-free matrix $\mathbf{P}^*$. As the loading matrix represents an orthogonal SVD basis, the sum of $\phi_k(l)$ over all $l$ components must equal 1. In the case of ideal decomposition, the distribution $\phi_k(l)$ is the Kronecker delta $\delta_{kl}$ while it should be smeared over many $l$ components if the decomposition is inaccurate.

It should be noted that in most cases, the first principal component differs drastically (in terms of the extraction accuracy) from the other ones because the first component consists of the mean data spectrum. To relax this difference we always subtract the mean spectrum from the data matrix $\mathbf{D}$ prior to the principal component decomposition. This operation is usually referred to as *data centering*.

Figure 3b shows the proximity between the eigenspectra and the true reference for the first ten components of the synthetic noisy dataset. From the comparison with an ideal case (Fig. 3a), it is evident that only the first obtained eigenspectrum reveals some proximity with the true spectrum, while the remaining eigenspectra show quite poor agreement with the noise-free references.

Figure 4 displays so-called *scree plots*—the variances of the extracted PCA components as a function of the component index. For the noise-free synthetic data (Fig. 4a), the ten meaningful components are visually well separated from the rest components. Note that, although noise was not introduced in this simulation, there were tiny random variations arising from the numerical inaccuracies of calculation, which contributed to a kind of numerical noise. However, even minor meaningful variations overpassed it, which ensured the easy visualization of the domain of meaningful components. The number of the phases (eleven) employed in the simulation fits perfectly the number of the observed meaningful components (ten). The difference (eleven versus ten) appears because the data centering procedure reduces the degree of freedom by one.

In contrast, the scree plot for the noisy dataset (Fig. 4b) indicates very poor correlations with the reference noise-free case. About 20 principal components can be roughly accounted as meaningful but there is no clear separation between them and the noise domain. The reconstruction with such a number of components leads to the unsatisfactory results dominated by noise and artifacts as will be shown in "Reconstruction of denoised datasets" section. We can conclude that PCA fails for the noisy synthetic dataset. It presumably fails also for the experimental dataset because its scree plot is quite similar to that for the noisy simulation.

The reason for the failure of classical PCA of noisy STEM EDX datasets is well known. PCA is able to extract meaningful variations only in the case when the superimposed noise has a similar level of variance in any given fragment of a dataset or, in other words, when the noise is *homoscedastic*. In fact, the dominant noise in XEDS spectra is Poisson noise, which is not homoscedastic.

### Weighted PCA
The Poisson noise in spectrum images can be converted into the homoscedastic one using a well-known property of Poisson noise, namely that its variance equals the mean count of the signal. Accordingly, the level of noise

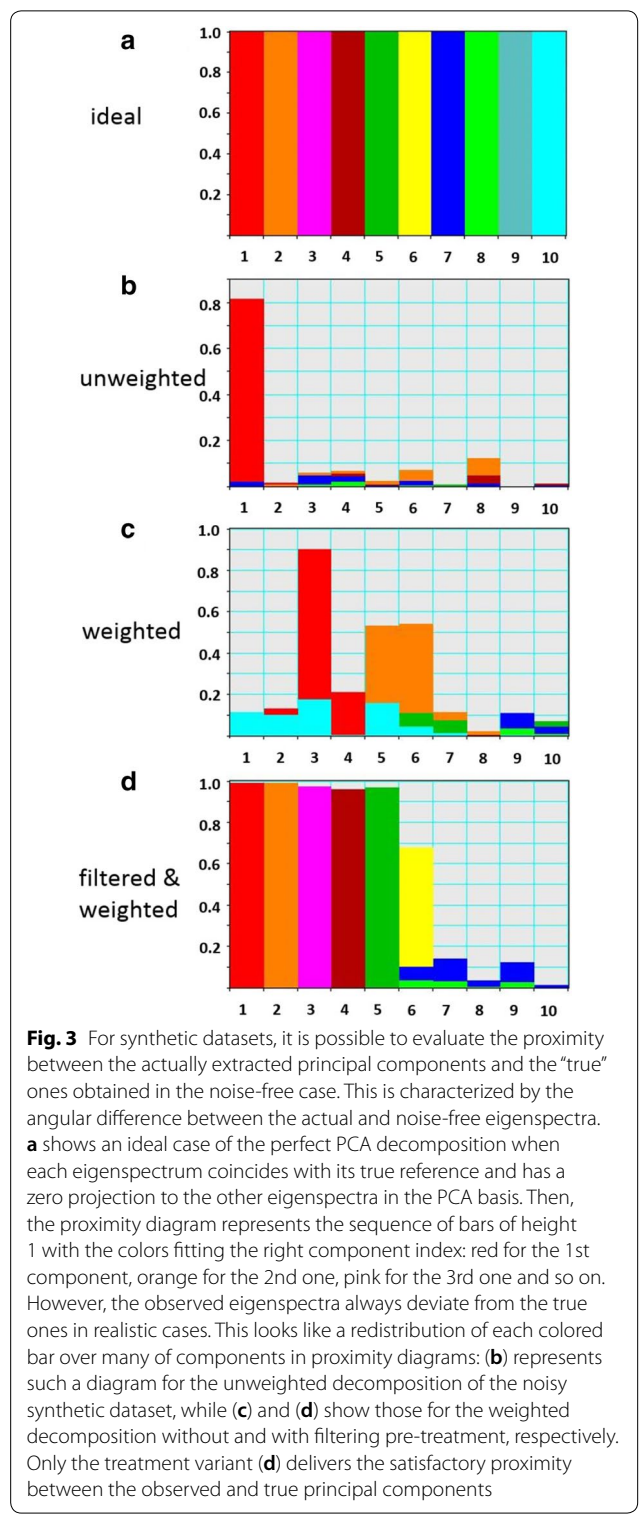

**Fig. 3** For synthetic datasets, it is possible to evaluate the proximity between the actually extracted principal components and the "true" ones obtained in the noise-free case. This is characterized by the angular difference between the actual and noise-free eigenspectra. **a** shows an ideal case of the perfect PCA decomposition when each eigenspectrum coincides with its true reference and has a zero projection to the other eigenspectra in the PCA basis. Then, the proximity diagram represents the sequence of bars of height 1 with the colors fitting the right component index: red for the 1st component, orange for the 2nd one, pink for the 3rd one and so on. However, the observed eigenspectra always deviate from the true ones in realistic cases. This looks like a redistribution of each colored bar over many of components in proximity diagrams: (**b**) represents such a diagram for the unweighted decomposition of the noisy synthetic dataset, while (**c**) and (**d**) show those for the weighted decomposition without and with filtering pre-treatment, respectively. Only the treatment variant (**d**) delivers the satisfactory proximity between the observed and true principal components

is proportional to the square root of the signal and can be equalized [27] by the following rescaling:

$$\widetilde{\mathbf{D}} = \mathbf{D} \div \mathbf{W} \qquad (4)$$

where symbol "÷" denotes the element-wise division and **W** is an $m \times n$ weighting matrix defined as

$$\mathbf{W} = (\mathbf{G} \otimes \mathbf{H})^{\frac{1}{2}} \qquad (5)$$

with **G** being an $m$-dimensional vector consisting of a mean image (image averaged over all energy channels) and **H** an $n$-dimensional vector consisting of a mean spectrum (spectra averaged over all pixels). The dyadic product $\mathbf{G} \otimes \mathbf{H} = \mathbf{G}\mathbf{H}^{\mathrm{T}}$ is expected to reflect the variation of the "true" signal in **D**, therefore the normalization to its square root should equalize the noise across a dataset.[1] In case the signal does not vary much from pixel to pixel, the weighting can be simplified by taking $\mathbf{G} = 1$. This kind of weighting will be referred in this paper to as *spectrum-only* weighting.

It should be stressed, however, that the elements of **W** provide only the estimates of the "true" signal level across the dataset. This estimation works typically quite well for STEM EELS but might be rather inaccurate in the case of STEM XEDS datasets as will be shown below.

After weighting the noise-free synthetic dataset, its scree plot (Fig. 4c) indicated 11 meaningful components, i.e. one more than that in the unweighted case. This can be explained by the non-linearity in the data variations, which was shown to increase the number of the observed components against the number of the latent factors [25]. Such non-linearity exists often in real-life objects and might be enhanced by the weighting rescaling.

Unfortunately, weighting does not improve the quality of principal component decomposition of the considered noisy synthetic dataset. Figure 3c demonstrates that the found eigenspectra still show poor correlation with the "true" ones. The 1st and 2nd true eigenspectra are partially retrieved in the 3rd–6th components of the noisy dataset but the rest meaningful components seem to be almost completely lost. In addition, the domains of meaningful and noise components in the scree plots (Fig. 4d) are not clearly separated for both noisy synthetic and the experimental datasets.

The failure of the weighting pre-treatment in STEM EDX spectrum-imaging has been already reported earlier [28, 29]. The reason for the problem is a high sparsity of typical STEM XEDS data that makes the evaluation of the elements of matrix **W** inaccurate. The sparsity of both the experimental and the noisy synthetic datasets in the present work was about 0.001, which means that only 0.1% of the elements in data matrix **D** were filled with a signal while 99.9% of them were empty. In this situation, the extracted mean spectrum and mean image suffer of

---

[1] In the case when weighting is combined with centering a data set, the former should be executed first because data centering destroys the basic properties of Poisson distribution.

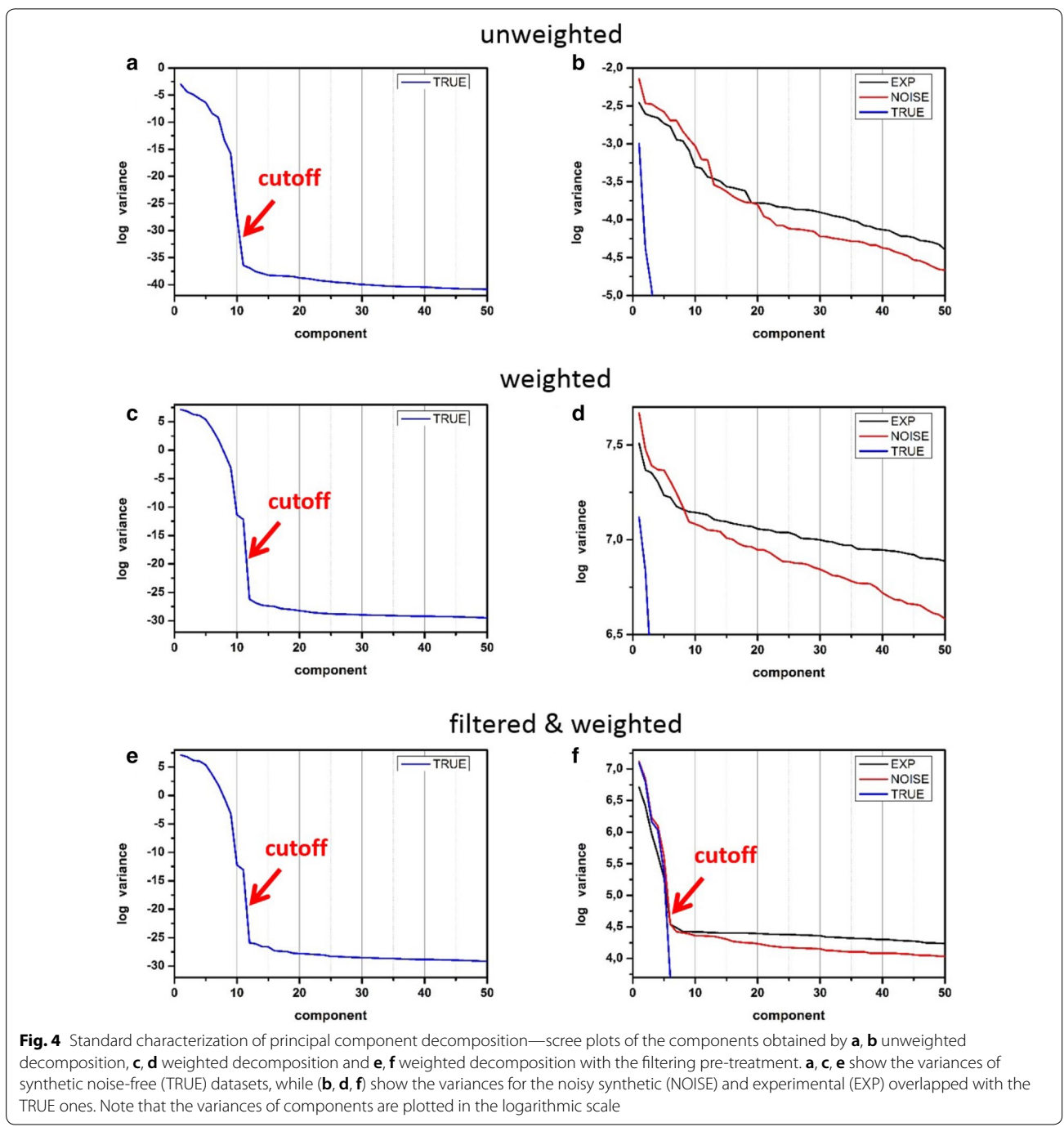

**Fig. 4** Standard characterization of principal component decomposition—scree plots of the components obtained by **a**, **b** unweighted decomposition, **c**, **d** weighted decomposition and **e**, **f** weighted decomposition with the filtering pre-treatment. **a**, **c**, **e** show the variances of synthetic noise-free (TRUE) datasets, while (**b**, **d**, **f**) show the variances for the noisy synthetic (NOISE) and experimental (EXP) overlapped with the TRUE ones. Note that the variances of components are plotted in the logarithmic scale

random variations that makes the weighting pre-treatment dangerous.

Appendix 2 considers the effect of sparsity on the weighing efficiency in details and Appendix 3 presents simulations confirming the conclusions of Appendix 2.

### PCA with smoothing filter pre-treatment
The most evident way to solve the sparsity issue outlined in the previous subsection is to smooth a dataset in either the spatial or energy directions prior the PCA treatment. The smoothing filtering would apparently reduce the sparsity of a dataset while hopefully preserving its general features. The simplest smoothing filter is binning the data as suggested by Kotula and van Bethlem [28]. Binning reduces also the size of a dataset that boosts the calculation speed and saves storage capacity. The disadvantage of intensive binning is a significant loss of the spatial and energy resolution, thus it might be employed only if the original dataset was oversampled for the required task (e.g. for obtaining an elemental map of given resolution or resolving certain features in spectra). Alternatively, data can be smoothed by Gaussian kernel filtering in the spatial or energy directions [29]. Gaussian smoothing fills the empty data elements even more efficiently than binning does while it deteriorates the resolution only slightly. On the other hand, Gaussian smoothing does not improve the calculation speed because the data size is unchanged. 2D Gaussian filtering in the spatial X- and Y-dimensions is most efficient in terms of reducing the data sparsity. Note that it must be performed before conversion of a data cube into matrix **D** because the spatial information is then retained more consequently.

In the present work, the combination of binning and Gaussian filtering was employed to overcome the sparsity issue. For the comparison purpose, the same filtering was applied to experimental, noisy and noise-free synthetic datasets. The datasets were first subjected to the $2 \times 2$ spatial binning, which provided a 4 times reduction in size. Then, the Gaussian kernel filtering with the standard deviation $\sigma = 1$ pixel was applied. To save the calculation time, the Gaussian function was truncated at 10% of its maximum such that the kernel mask included 12 pixels around the central pixel (see [29] for details). No smoothing in the energy direction was applied.

The filtering pre-treatment dramatically improves the quality of principal component decomposition as demonstrated in Fig. 3d. The eigenspectra of at least 6 major components of the noisy synthetic dataset are now in a good proximity with the reference noise-free eigenspectra. The 7th component (blue in Fig. 3c) lies at the limit of detectability—although the proximity function is rather wide-spread, its maximum seems to stay at the approximately correct position.

The scree plot of the noisy synthetic dataset in Fig. 4f now clearly visualizes two domains - the domain of the meaningful components with a higher variance and the noise domain where the variance follows a steadily decreasing line. The border between two domains is located near the 6–8th component. Superposing the scree plots of the noisy and noise-free datasets reveals that they closely follow each other up to the 6th component. On the other hand, the scree plot of the experimental dataset is very similar to that of the noisy synthetic one, which suggests that most of components of the real-life object are retrieved accurately.

## Truncation of principal components and reconstruction

At the next step of the PCA treatment, it is assumed that the only few major PCA components carry the useful information while the remaining minor components represent noise. Therefore, a dataset can be reconstructed using only $k$ ($k \ll n$) major components as illustrated in Fig. 1. This truncation implies a reduction of the effective dimensionality of data from $n$ to $k$ in the energy dimension. Accordingly, a dataset is significantly denoised because most of the noise variations are removed with the omitted minor components.

The key question of the PCA analysis is *how to truncate?* Retaining too many components would marginalize denoising while retaining too few components might distort the meaningful variations. A naive consideration suggests that the number of retained components should correspond to the number of latent factors $L$ governing the spectral changes in data. The latter, however, is often not known even approximately. Furthermore, the number of components retrieved (or potentially retrievable) from PCA can significantly differ from $L$ because:

- It can be higher than $L$ as a result of experimental artifacts like changing the beam current in the course of scanning or changing response of the detector.
- It can be higher than $L$ due to non-linearities in the spectra formation such as absorption in the mixture of light and heavy phases. These non-linearities manifest themself as the appearance of additional dimensions in the energy space unrelated with any latent factor [25].
- It can be smaller than $L$ if the variance of some minor components approaches the variance of noise. Then, these components might be irretrievable from PCA [30–32].

The latter effect is extremely important for practical cases and will be discussed on the example of the considered object in the next subsection.

### Loss of minor principal components in noisy datasets

Within the framework of a so-called *spiked covariance model*, Nadler [33] has shown that PCA is able to retrieve a meaningful component only if the following inequality is satisfied:

$$\frac{m}{n} \geq \left( \frac{\sigma^2}{\lambda^*} \right)^2 \tag{6}$$

where $m$ and $n$ are, as above, the number of STEM pixels and energy channels, respectively, $\sigma^2$ is the variance of homoscedastic noise and $\lambda^*$ is the "true" (not corrupted

**Table 2 Extracted variances ($\lambda$) and "true" variances ($\lambda^*$) of the noisy and true synthetic dataset**

| Component | $\lambda$ | $\lambda^*$ | $\frac{\lambda^*}{\sigma^2}$ | Retrievable |
|---|---|---|---|---|
| 1 | 1228 | 1214 | 43.2 | ✔ |
| 2 | 938.3 | 906.4 | 32.3 | ✔ |
| 3 | 509.2 | 482.2 | 17.2 | ✔ |
| 4 | 444.7 | 422.8 | 15.1 | ✔ |
| 5 | 273.8 | 214.6 | 7.65 | ✔ |
| 6 | 94.04 | 40.05 | 1.43 | ✔ |
| 7 | 83.53 | 6.571 | 0.234 | ✔ |
| 8 | 81.98 | 0.5804 | 0.0207 | – |
| 9 | 80.61 | 0.04119 | 1.47e−3 | – |
| 10 | 78.30 | 5.13e−6 | 2.01e−7 | – |
| 11 | 78.28 | 1.63e−6 | 6.43e−8 | – |

The level of homoscedastic noise $\sigma^2$ is 28.07. According the Nadler model [33], a component is retrievable if the value in the 4th column exceeds $\sqrt{\frac{n}{m}}$, which is 0.245 for the number of channels $n = 1200$ and the number of pixels $m = 19920$ by noise) variance of a meaningful component. If $\lambda^*$ appears to be smaller than $\sqrt{\frac{n}{m}}\,\sigma^2$, a component cannot be retrieved even theoretically.

The synthetic datasets presented in this paper provide a good possibility to test this prediction as the noise variance and meaningful noise-free variances of all components are precisely known. Table 2 demonstrates that about 7 meaningful components can be theoretically retrieved in the course of principal components decomposition. This is in good agreement with the results in Fig. 3d, which suggest that the 7th component appears at the limit of detectability.

Formula (6) also predicts that the range of detectable components can be extended by the smoothing filter pre-treatment. Indeed, filtering reduces $\sigma^2$ thus smaller $\lambda^*$ can be retrieved[2] [29]. This provides an additional argument in favor of filtering pre-treatment described in "PCA with smoothing filter pre-treatment" section.

An estimation within the spiked covariance model is instructive for understanding why minor components might be lost in the course of PCA. However, Eq. (6) is based on the precise knowledge of "true" eigenvalues $\lambda^*$ that are not directly accessible in the experiment. In the next subsections we consider existing practical truncation methods that do not require the knowledge of these parameters.

### Scree plot method for truncation of principal components

Historically, one of the earliest and most popular methods is analyzing a scree plot. This is based on the assumption that meaningful components show a data variance noticeably higher than that of the noise. The variance of noise components is assumed to follow some smooth curve, thus the meaningful and the noise domains can be visually separated on scree plots such as those in Fig. 4f.

In most cases, the scree plot analysis leads to satisfactory results. However, this kind of truncation requires manual evaluation and is not accurately reproducible as different persons tend to set the border between the visually distinct regions slightly differently. For the considered noisy synthetic and experimental datasets in Fig. 4f, the border can be subjectively set between 6 and 8. It is also quite difficult to incorporate the scree plot approach into automatic algorithms because the behavior of the noise variance might vary significantly and the factorization of its dependence in the noise domain is problematic.

### Analytical model-based methods for truncating principal components

Recently, several approaches for analytical determination of the optimal number of principal components have emerged [34–37]. These methods are based on certain models for the mixture of noise and useful signal typically assuming the Gaussian nature of noise.

A representative example for this family of truncation methods is an approach by Gavish and Donoho [36], who considered a spiked covariance model with a known level of homoscedastic noise $\sigma$ and searched for the optimal threshold minimizing the mean squared error between the reconstructed and noise-free datasets. They found that eigenvalues $\lambda$ of the retained components should satisfy[3]:

$$\lambda \geq \frac{n}{m}\left(\alpha\left(\frac{m}{n}\right)\right)^2 \sigma^2 \qquad (7)$$

where, as before, $m$ is the number of pixels and $n$ is the number of energy channels and $\alpha\left(\frac{m}{n}\right)$ is a tabulated parameter dependent on $\frac{m}{n}$. In the simplified case when $m = n$, (7) is reduced to $\lambda \geq \frac{16}{3}\sigma^2$. Note that Eq. (7) consists of observable eigenvalues $\lambda$ (not "true" eigenvalues $\lambda^*$ as in 6), and therefore, can be used for practical truncation of principal components.

The approach of Gavish and Donoho as well as other similar approaches require the precise knowledge of the level of homoscedastic noise $\sigma$ that is, in practice, very difficult to extract from experimental data (see [35] for details). This can be performed by subtracting all meaningful principal components and evaluating the retained

---

[2] The more accurate analysis requires an introduction of the effective number of independent pixels $m_e$, which can be also affected by filtering [29].

[3] We rewrote original formula (3) from [36] to fit the definitions and notations used in the present paper.

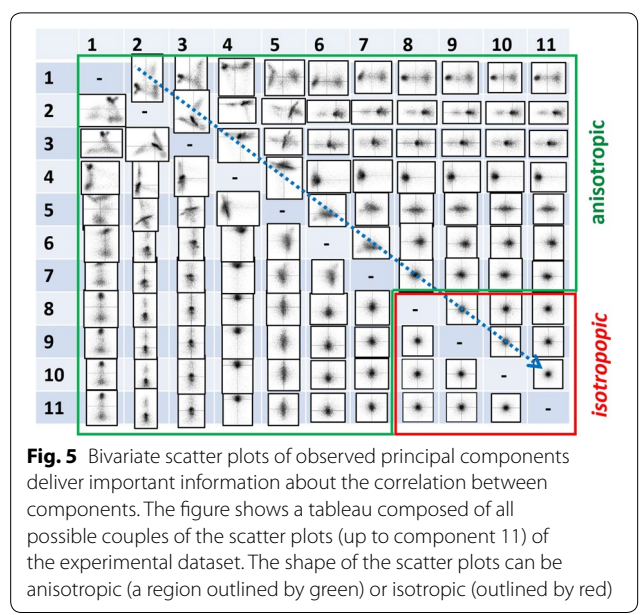

**Fig. 5** Bivariate scatter plots of observed principal components deliver important information about the correlation between components. The figure shows a tableau composed of all possible couples of the scatter plots (up to component 11) of the experimental dataset. The shape of the scatter plots can be anisotropic (a region outlined by green) or isotropic (outlined by red)

data fraction with the so-called *real error function* [38]. The approach implies the time-consuming iterative evaluation of the noise level alternated with the cut-off of meaningful components. Furthermore, as will be demonstrated in "Comparison of different truncation methods" section and Appendix 4, the accuracy of the resulted truncation is not guaranteed in the case of STEM EDXS spectrum-imaging.

### Anisotropy method for truncation of principal components

To overcome the limitations highlighted in the previous subsections, we suggest a novel practical method for truncation of principal components, which is flexible, objective and can be easily implemented in automatic algorithms.

Consider a *scatter plot*—a joint plot of scores for two given principal components. A number of examples for scatter plots of the analyzed object are presented in Appendix 3. These plots are quite informative because their geometry immediately reveals correlation or absence of correlation between given components. Figure 5 shows a tableau composed of all possible scatter plots for the principal component decomposition of the experimental dataset up to component 11. First of all, note that this tableau is symmetric relative to its diagonal as reversing the indexes in a couple is equivalent to mirroring a corresponding scatter plot. Second, the tableau can be separated into two regions: one (green outlined) with the quite asymmetric scatter plots and another (red outlined) with the scatter plots of round shape. The former region (green) consists of couples of components with at least one meaningful component. If the other component is noise, the scatter plot contains a spreading of the meaningful distribution along either vertical or horizontal axis. In the case when both components are meaningful, the resulting scatter plot is more complicated and asymmetric. The latter region (red) consists of couples of compounds both representing pure noise. As random noise is uniform in all energy directions, the distribution of their scores is expected to be round. In the most common case of the Gaussian nature of noise, such scatter plots should represent the bivariate Gaussian distribution.

The described property of scatter plots can be employed as an objective criterion to discriminate meaningful and noise components. The easiest way to find the border between two regions is tracking the sequential couples of components 1–2, 2–3, 3–4... as shown by the dashed arrow in Fig. 5. Automatic truncation technique implies now a method to quantify the anisotropy of a given scatter plot. We will consider several criteria directly or indirectly related with the factorization of a bivariate distribution. One of them is the *covariance* (Cov) of two distributions (scores of two components in this case):

$$\text{Cov}(T_1, T_2) = \frac{1}{m} \sum_i^m (T_1(i)T_2(i)) \qquad (8)$$

where $T_1$ and $T_2$ are the scores of two given principal components constituting the plot and $m$ is the number of pixels. Computation of the covariance in Eq. (8) is simplified by the fact that the mean values of $T_1$ and $T_2$ are zero for the centered principal components decomposition.

Another way to evaluate the anisotropy could be calculating a third-order momentum of a scatter plot. As an example of such approaches, we calculate the bivariate *skewness* (Ske) defined after Mardia [39] as

$$\text{Ske}(T_1, T_2) = \frac{1}{m^2} \sum_{i,j}^m \left[ \frac{T_1(i)T_1(j) + T_2(i)T_2(j)}{\text{Cov}(T_1, T_2)} \right]^3 \qquad (9)$$

Alternatively, we investigated a group of methods that utilize the geometry of the bivariate score distribution. One can digitize the joint distribution of $T_1$ and $T_2$ in a kind of a two-dimensional $t \times t$ grid $S$ where each cell of the grid corresponds to certain small ranges of the $T_1$ and $T_2$ values. The $s_{lq}$ cells are assigned to zero if none of the $(T_1(l), T_2(q))$ couples appear in the given ranges. Otherwise $s_{lq}$ cells are assigned to the number of events with both $T_1$ and $T_2$ satisfying the corresponding ranges. In fact, this digitization is performed every time when a display of a scatter plot is built (see examples of displayed scatter plots in Fig. 5 and in Appendix 3). Based on this digitized representation we may now analyze, whether the bivariate score distribution is a dyadic product of two independent distributions or not. To this end we compute the following criterion

$$\text{Pur}(T_1, T_2) = \sum_{l,q=0}^{t} \left[ \frac{s_{lq}}{\text{tr}(S)} \right]^2 \tag{10}$$

which is 1 for the ideal dyad (no correlation) while it becomes larger for general matrices (correlated case). Note that a similar task—the discrimination between pure (i.e., dyad) and mixed quantum states—is known in quantum mechanics. There one computes the so-called *purity* of a normalized density matrix in a similar way and the resulted value is 1 for a pure state and less than 1 for a mixed state. The anisotropy criterium introduced in Eq. (10) roughly amounts to the inverse of the matrix purity in quantum mechanics. This inversion in definition is made to match the tendencies in other anisotropy criteria.

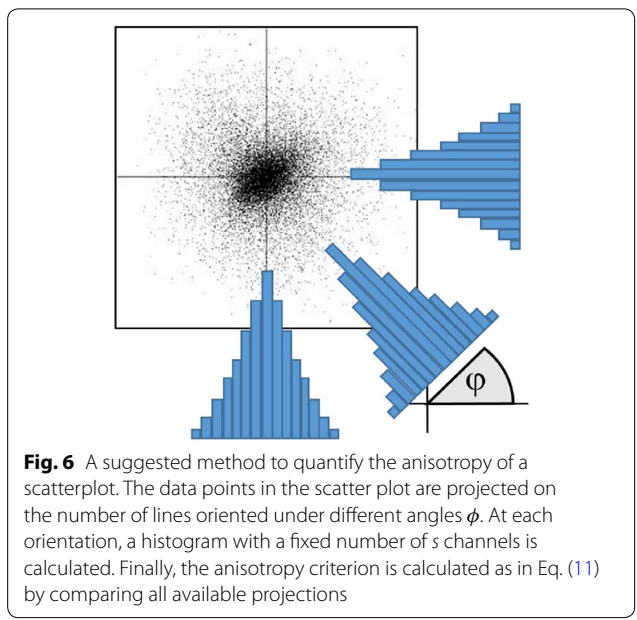

**Fig. 6** A suggested method to quantify the anisotropy of a scatterplot. The data points in the scatter plot are projected on the number of lines oriented under different angles $\phi$. At each orientation, a histogram with a fixed number of $s$ channels is calculated. Finally, the anisotropy criterion is calculated as in Eq. (11) by comparing all available projections

Finally, we explored the simplest method that analyses the histograms of grid $S$ under the different projections as sketched in Fig. 6. The anisotropy of the plot can be then evaluated as:

$$\text{His}(T_1, T_2) = \frac{1}{ps} \sum_{l=1}^{s} \sum_{\varphi=0}^{\pi/2} \frac{(H(l,\varphi) - \bar{H}(l))^2}{\bar{H}(l)} - 1 \tag{11}$$

where $H(l,\varphi)$ is a histogram of the projection under angle $\varphi$ and $\bar{H}(l)$ is the average over all projections. Here $p$ is a number of projections in the range of $\varphi = -\pi/2 \ldots \pi/2$ and $s$ is a number of channels in a histogram. In the case of a perfectly isotropic scatter plot, there are only random deviations of $H(l,\varphi)$ from rotational average $\bar{H}(l)$ following the Poisson distribution. This means that their variance must equal their average value. Therefore, the expression under the sum should approach 1 and the whole expression averaged over all projections and channels should approach zero for a perfectly round scatter plot.

Figure 7 compares the different quantification methods for the series of the scatter plots of the experimental dataset. The covariance and matrix purity criteria oscillate too much for high-index components, which does not allow to separate the meaningful and noise domains reliably. The skewness criterion might approach quite low values in the noise domain but sporadic heavy outliers make this method unreliable as well. Additionally, the calculation of skewness requires an order of magnitude longer computation time than the other methods. After all, the only method of projected histograms (Fig. 7e) provides a clear separation between the anisotropic and isotropic domains. Zooming into the transient region (inset in Fig. 7e) reveals that this criterion oscillates very close to zero in the noise domain and any departures from isotropy are evident there. A small bias in the positive direction might appear as the variances of neighboring noise components are slightly different. This can be canceled by normalizing the scores of both components to their variance.

It should be, however, stressed, that the anisotropy method fails for sparse STEM XEDS data. In this case, the anisotropy criterion shows quite high values both for the meaningful and noise components. The reason for that is apparent—if only a few data elements are assigned to one and the rest are zeros, a *random-like* asymmetry of scatter plots might be observed even if the underlying data distribution is isotropic. Therefore, a treatment reducing the data sparseness like that described in "PCA with smoothing filter pre-treatment" section is obligatory prior application of the anisotropy method.

### Comparison of different truncation methods

Table 3 lists the number of principal components to truncate according to the methods described in the previous subsections. The scree plot and anisotropy methods suggest similar truncation cut-offs, which are in agreement with the theoretical prediction of Eq. (6). The analytical approach by Gavish and Donoho suggests the same cut-off in the case of the experimental dataset but overestimates dramatically the number of meaningful components for the synthetic noisy one. Another examples in Appendix 4 demonstrate that this is not a singular exception but rather an indication of instability of the method when applied to STEM XEDS spectrum images.

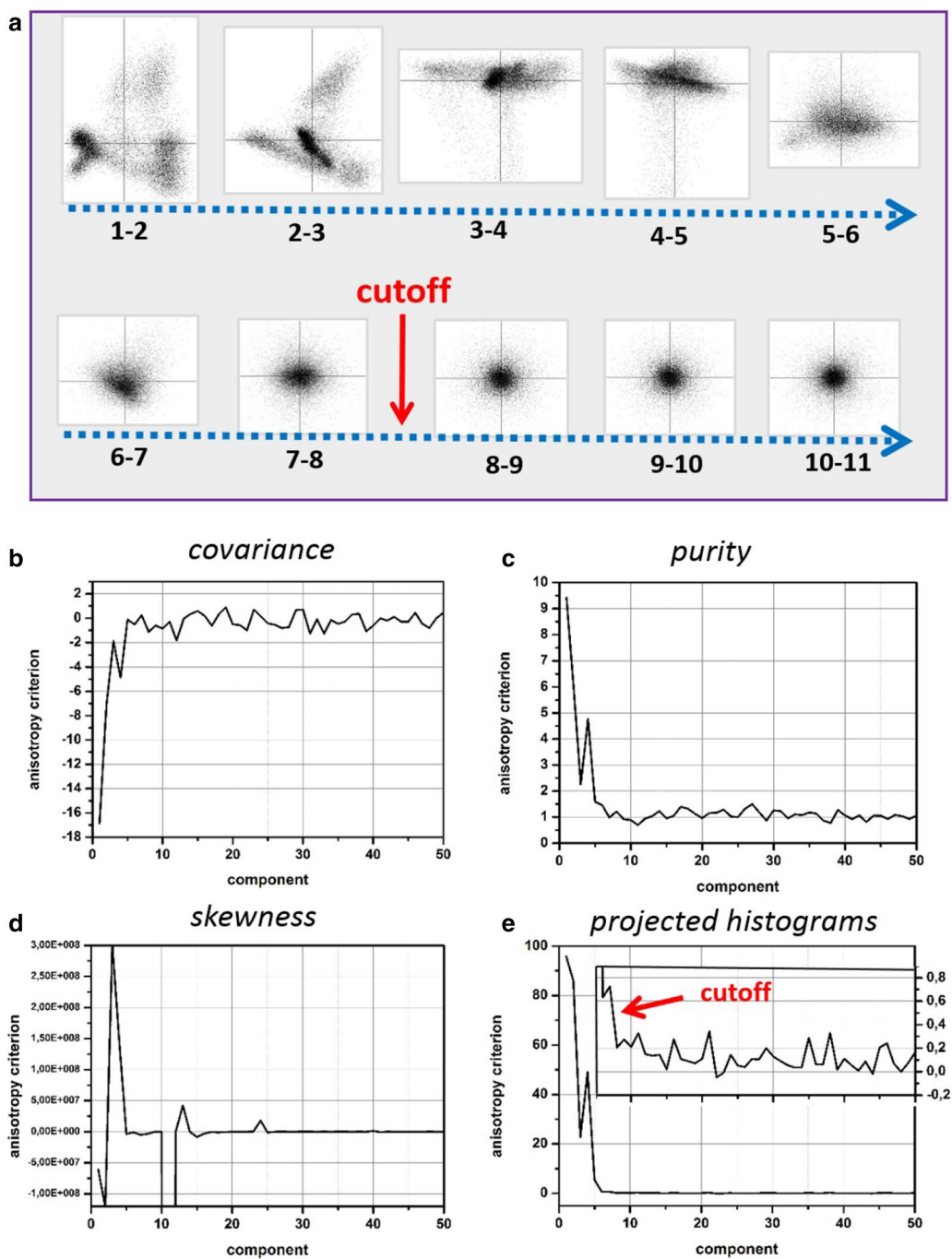

**Fig. 7** **a** The easiest way to localize the border between the anisotropic and isotropic regions in Fig. 5 is to track the sequential couples of scatter plots along the blue-dashed arrow while analyzing their anisotropy as shown in (**a**). The point where the scatter plot turns from anisotropic to isotropic denotes a reasonable cut-off for the meaningful principal components. To make the procedure automatic the quantitative criterion for anisotropy is needed; **b**–**e** compare the different anisotropy criteria: covariance (**b**), multivariate skewness (**c**), matrix pureness (**d**) and projected histograms (**e**). The component index plotted in the horizontal axis corresponds to the lowest index of the evaluated couple of components. Among all the considered methods, the method of projected histograms (**e**) performs best in separating the anisotropic and isotropic regions. The anisotropy criterion oscillates very closely to zero in the isotropic region as apparent from the inset in (**e**)

**Table 3** The number of components to truncate according to the different truncation methods: the evaluation of a scree plot with visual localisation of the inflection point ("Scree plot method for truncation of principal components" section), the approach of Gavish and Donoho ("Analytical model-based methods for truncating principal components" section) and the anisotropy method ("Anisotropy method for truncation of principal components" section) with using the projected histograms and the anisotropy threshold of 0.5

| Dataset | Scree plot | Gavish and Donoho | Anisotropy |
|---|---|---|---|
| Synthetic | 6–8 | 30 | 7 |
| Experimental | 6–8 | 7 | 7 |

Although the scree plot and anisotropy methods perform similarly, the latter offers a crucial advantage—the cut-off can be determined less subjectively. Localizing the inflection point in a scree plot is straightforward but might require a number of tunable parameters in an unsupervised treatment. In contrast, the method of scatter plot anisotropy can be easily incorporated into an automatic algorithm. The anisotropy oscillates around zero in the noise domain, which is very beneficial compared to a scree plot, where the variance decays slowly from an unknown level. Therefore, a single threshold parameter can be used to discriminate the meaningful and noise domains. This parameter represents the relative deviation from isotropy that can be still tolerated. To our experience the threshold parameter can be set to 0.5–1.0 for the case of STEM XEDS and EELS spectrum-imaging depending on the required accuracy of the detection of minor principal components. It is also possible to define the threshold adaptively depending on the measured variation of anisotropy in the region with very high indexes of components.

The suggested anisotropy method employs a very basic property of random noise—its directional isotropy. It does not put any strict assumptions on the specific nature of noise—Poissonian, Gaussian, or mixed. The synthetic data presented in this paper are corrupted by the Poisson noise, which is converted to the Gaussian-like one after the correctly performed weighting procedure. In real experimental data, some small fractions of noise might come from the imperfections of registration that makes the noise distribution more complicated. Some hints for that are the different slopes of the scree plots in the noise domains for experimental and syntetic datasets in Fig. 4f. Nevertheless, the anisotropy method delivers identical truncation cut-offs for both the datasets, which suggests a certain robustness against the nature of noise.

Appendix 4 shows more examples of application of the anisotropy method for truncating principal components in STEM XEDS data. The anisotropy criterion behaves similarly (compare Figs. 7e, 13b and 14b) in the variety of STEM XEDS data—it shows quite high values for the meaningful components and then oscillates around zero in the noise domain. Furthermore, it has been demonstrated that the method works reliably for STEM EELS spectrum-images as well [40].

### Reconstruction of denoised datasets

After truncating principal components, a dataset can be reconstructed as

$$\mathbf{D} \approx \left[ \tilde{\mathbf{T}} \tilde{\mathbf{P}}^{\mathrm{T}} \right]_k \cdot \mathbf{W} \tag{12}$$

where the index $k$ means that the energy dimension $n$ was reduced to $k$ in both weighted score $\tilde{\mathbf{T}}$ and loading $\tilde{\mathbf{P}}$ matrices as illustrated in Fig. 2. Here, the symbol $\cdot$ means the element-wise multiplication on weighting matrix $\mathbf{W}$. In the case when the principal component decomposition and truncation were performed reasonably, the reconstructed data matrix incorporates all meaningful data variations but noise. The conversion into the original data arrangement (line scan, data cube or cube of higher dimension) results in a new spectrum-image greatly denoised compared with the initial one.

Data can be then subjected to the further processing depending on the treatment goals. For instance, Fig. 8 shows elemental maps extracted from the noisy synthetic dataset by integration of the corresponding XEDS lines. The maps for the best PCA treatment strategy (filtering and weighting, 7 components-truncation) appear to be noticeable improved compared to those from the raw spectrum image. Note that the effect is *not* due to a trivial improvement of the data appearance by filtering (compare Fig. 8b, e). The non-optimal PCA strategies might eventually denoise elemental maps as well, but they often result in dangerous artifacts. For instance, the Hf layer disappears and manifests itself as a false Ta layer in Fig. 8c, d.

The corresponding elemental maps from the experimental spectrum-image are shown in Fig. 9. The comparison between Figs. 8 and 9. confirms that the noisy synthetic and experimental data sets behave very similarly under the various strategies of the PCA treatment. Therefore, the trends discovered by analyzing the synthetic data can be reliably extended towards experimental data of real-life objects.

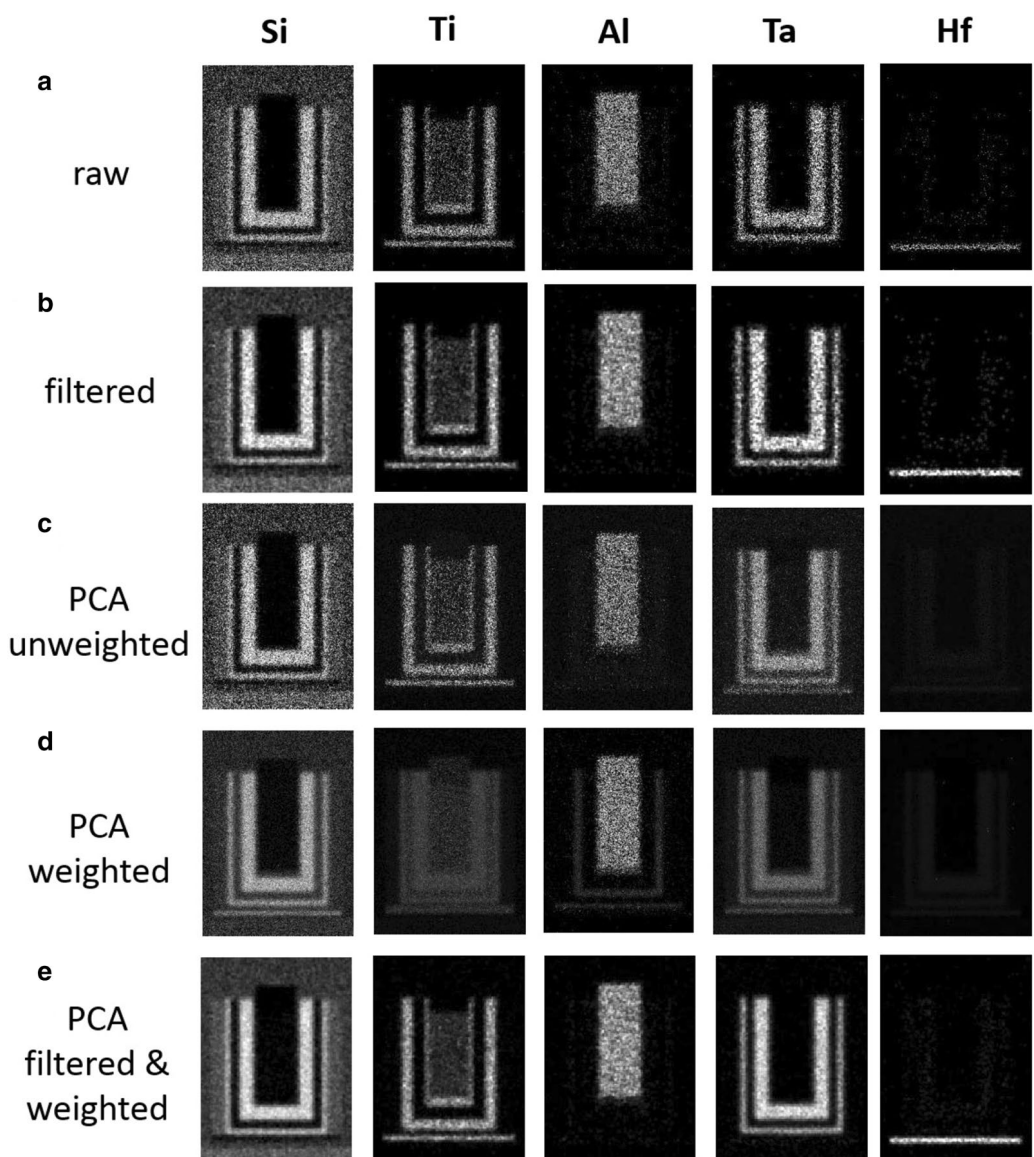

**Fig. 8** Elemental maps extracted from (**a**) raw and (**b**) filtered noisy synthetic datasets; **c–e** represent elemental maps extracted from the reconstructed datasets according to the (**c**) unweighted PCA, (**d**) weighted PCA and (**e**) filtered and weighted PCA treatments. The number of principal components used in reconstruction was 22 for (**c**), 9 for (**d**) and 7 for (**e**). For fair comparison, all maps are displayed with identical contrast settings. The PCA treatment improves noticeably the visibility of the chemical components, especially the weak-signal elements—Ta and Hf. Note that the treatment variants (**c**) and (**d**) cause the artificial intermixing of the Ta and Hf layers

## Conclusions

We conclude that experimental STEM XEDS spectrum-images acquired with modern STEM instrumentation and typical acquisition settings can be noticeably denoised by application of PCA. Here, two pre-treatments of the typically sparse STEM XEDS datasets are ultimately needed for the successful PCA: smoothing and weighting.

A crucial step for denoising spectrum images is the truncation of principal components. Theoretical consideration shows that the optimal number of retained components depends on the ratio between the levels of noise and expected meaningful variations in an object as well as on the number of pixels in a spectrum image.

We presented a promising method for optimally truncating principal components based on the analysis of the anisotropy of scatter plots resulting from the principal components decomposition. This method can be easily

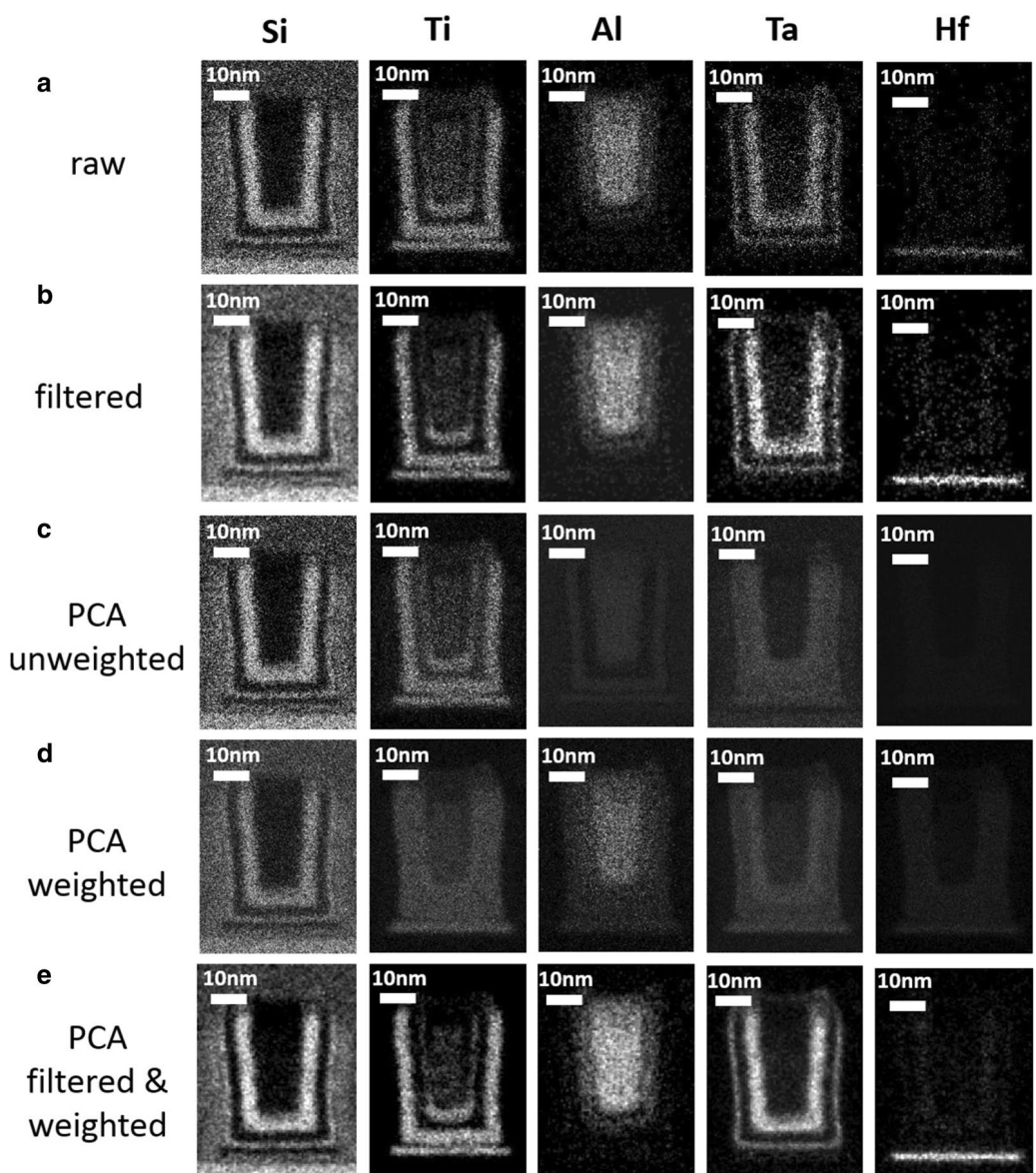

**Fig. 9** Elemental maps extracted from (**a**) raw and (**b**) filtered experimental datasets; **c**–**e** represent elemental maps extracted from the reconstructed datasets according the (**c**) unweighted PCA, (**d**) weighted PCA and (**e**) filtered and weighted PCA treatments. The number of principal components used in reconstruction was 18 for (**c**), 12 for (**d**) and 7 for (**e**). For fair comparison, all maps are displayed with identical contrast settings. The PCA treatment improves noticeably the visibility of the chemical components, especially the weak-signal elements—Ta and Hf. Note that the treatment variants (**c**) and (**d**) cause the artificial intermixing of the Ta and Hf layers

implemented in automatic algorithms, which promotes a smooth, unsupervised workflow.

Given the straightforward implementation of the presented PCA workflow and the power of the method for denoising datasets containing, e.g., only small concentrations of elements with sparse spectra, we anticipate a further increase in PCA applications to STEM-based spectrum-images as well as other hyperspectral techniques with similar dataset properties.

## Methods

### Experimental details

The STEM XEDS spectrum-imaging was performed in the Titan G2 (S)TEM microscope operating at 300kV and equipped with the 4-windows SDD XEDS detector. The TEM cross section of the CMOS device was prepared by FIB at 30 kV followed by Ga ion milling at 5 kV. The final thickness of the sample was approximately 50 nm.

The STEM scanning with collecting the XEDS signal was executed within 6 minutes in the multi-frame mode across the 244 × 336 pixel rectangle covering the area of approximately 40 × 50 nanometers. The probe size was about 0.2 nm and the beam current was 120 pA. Although the spectra were originally acquired with 4096 energy channels, the data cube was then truncated to 1200 channels in the range of 0.2–12.2 keV that covered all useful XEDS peaks.

### Details of simulation

A phantom object that mimicked a real CMOS transistor was generated as shown in Fig. 2b. The geometry of the layers was greatly simplified but their volume fractions were reproduced reasonably accurate. The composition of each layer was set according Table 1 and then the borders among them were numerically smeared out to mimic the roughness of the layers in the real device and the spread of the STEM probe along the 50 nm sample thickness.

XEDS spectra were generated using the simulation program DTSA-II [41] developed in National Institute of Standards and Technology. The simulation employed an acceleration voltage of 300 kV, a realistic model for an SDD detector, a sample thickness of 50 nm and the compositions of the layers as listed in Table 1. The generated spectrum-images consisted of the same number of STEM pixels (244 × 336) and energy channels (1200) as the experimental dataset.

The synthetic data were prepared in two variants: one with no noise (the counts were represented by floating numbers, not truncated to integers) and another with a Poissonian noise added according to the nominal signal at each data point (here the counts were represented by integers as appearing in the experimental set). For the best compliance with the experiment, the synthetic spectrum-images were scaled such that the total number of counts in the range of 0.5–12 kV coincided with that of the experimental dataset.

### Additional files

> **Additional file 1.** Simulated STEM XEDS spectrum-image of a phantom CMOS device with adding Poisson noise.
>
> **Additional file 2.** Simulated STEM XEDS spectrum-image of a phantom CMOS device without adding noise.

### Abbreviations

TEM: transmission electron microscopy; STEM: scanning transmission electron microscopy; XEDS: X-ray energy-dispersive spectroscopy; EELS: electron energy-loss spectroscopy; SDD: silicon drift detector; PCA: principal component analysis; SVD: singular value decomposition; CMOS: complementary metal-oxide semiconductor; HAADF: high-angle annular dark field; ToF-SIMS: time-of-flight secondary ion mass spectroscopy.

### Authors' contributions
PP collected the experimental data, developed the evaluation strategy and the code. AL developed the various methods of truncating principal components and contributed to writing the manuscript. Both authors read and approved the final manuscript.

### Author details
[1] Department of Physics, Technical University of Dresden, Dresden, Germany. [2] Leibniz Institute for Solid State and Materials Research (IFW), Dresden, Germany.

### Acknowledgements
Nicholas W. M. Ritchie, Natinal Institute of Standards provided the useful comments on adaptation of the DTSA-II package to TEM simulations.

### Competing interests
The authors declare that they have no competing interests.

### Availability of data and materials
The PCA treatment of the experimental and simulated objects was performed with the *temDM MSA* package downloadable at http://temdm.com/web/msa/. The synthetic datasets described in the present paper can be found at the same website under the name "synthetic STEM XEDS spectrum-images of a CMOS device".

### Funding
The authors acknowledge funding from Deutsche Forschungsgemeinschaft "Zukunftskonzept" (F-003661-553-Ü6a-1020605) and from European Research Council under the Horizon 2020 program (Grant 715620). The support by the Open Access Publishing Funds of the SLUB/TU Dresden is acknowledged.

## Appendices

### Appendix 1: Effect of data sparsity on weighting XEDS spectrum-images

Consider a simplified spectrum-only weighting where the elements of an $m \times n$ data matrix $\mathbf{D}$ are normalized to the square root of a mean spectrum. In other words, all columns of $\mathbf{D}$ are normalized to the square root of the mean value $h_j$ over the corresponding $j$-th column of $\mathbf{D}$. Assuming random noise, the data variance $\nu_j^2$ in each column can be separated into "true" variance $\mu_j^2$ and noise variance $\sigma_j^2$ as:

$$\nu_j^2 = \mu_j^2 + \sigma_j^2 \tag{13}$$

with the "true" variance:

$$\mu_j^2 = \mathrm{Var}\left(\frac{d_{ij}^*}{\sqrt{h_j}}\right) = \frac{\mathrm{Var}(d_{ij}^*)}{|h_j|} \tag{14}$$

and the noise variance:

$$\sigma_j^2 = \mathrm{Var}\left(\frac{d_{ij} - d_{ij}^*}{\sqrt{h_j}}\right) = \frac{\mathrm{Var}(d_{ij} - d_{ij}^*)}{|h_j|} \tag{15}$$

where $d_{ij}$ denote the actual elements of data matrix $\mathbf{D}$ while $d_{ij}^*$ refer to "true", noise-free values. The variance of Poissonian noise in each column is equal to the "true" data mean $h_j^*$ along this column. In the case the sufficient

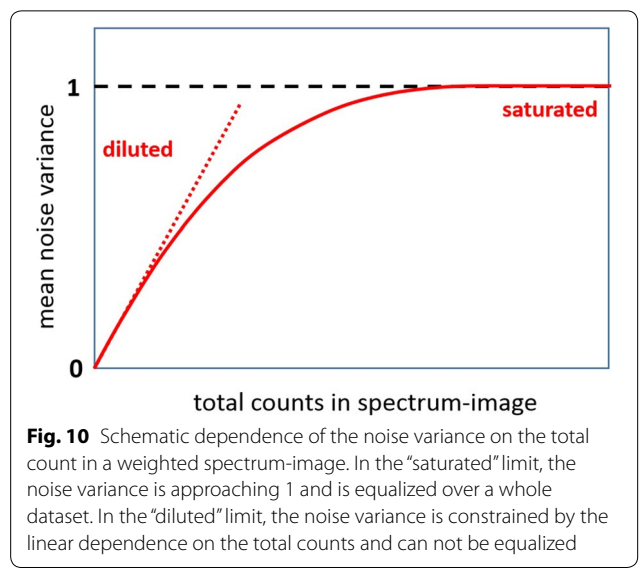

**Fig. 10** Schematic dependence of the noise variance on the total count in a weighted spectrum-image. In the "saturated" limit, the noise variance is approaching 1 and is equalized over a whole dataset. In the "diluted" limit, the noise variance is constrained by the linear dependence on the total counts and can not be equalized

signal is present, $h_j \approx h_j^*$ and $\sigma_j^2 \to 1$, therefore the noise is equalized among all columns. The situation is quite different if a dataset is sparse. Consider a limiting case where each column consists of no more than one XEDS count. A lot of columns will contain of entirely zeros and $h_j$ will be also zero. To avoid divergence, such columns must be simply set to zero in the course of the weighting procedure. The columns consisting of one count should be normalized to $\sqrt{h_j} = \frac{1}{\sqrt{m}}$. The data variance in each column is

$$v_j^2 = \begin{cases} \frac{(\sqrt{m} - \frac{\sqrt{m}}{m})^2}{m} \approx 1 & \text{one count in column} \\ 0 & \text{no counts in column} \end{cases} \quad (16)$$

oscillating between two extreme numbers, 1 and 0. Note that the data variance $v^2$ averaged over all columns is now proportional to the total number of counts $c$ in a spectrum image:

$$v^2 = \frac{c}{n} \quad (17)$$

as sketched in Fig. 10. The total data variance is greatly reduced due to the presence of a number of data elements $d_{ij}$ with zero values. The noise variance is constrained as

$$\sigma^2 \le \frac{c}{n} \quad (18)$$

and apparently can not be normalized to 1 in the "diluted" limit.

In the situations intermediate between the "diluted" and "saturated" limits, the average data variance (and therefore the noise variance) must be lower than 1 as shown in Fig. 10. That means that the noise can not be scaled to the fixed value 1 but varies instead in each column depending on actual $h_j$, which takes more or less randomly the small discrete values $0, \frac{1}{m}, \frac{2}{m} \ldots$ depending on how much counts are observed in a column. Such *digitization* of the data variance in the course of the weighting procedure is the direct consequence of the sparseness of the typical STEM XEDS signal. The issue does not exist in STEM EELS data that provide always the sufficient signal for accurate evaluation of the mean spectrum values.

These speculations can be extended to the case when weighting employs the normalization to both mean spectrum and mean image. The issue occurs if any of elements of weighting matrix **W** introduced in Eq. (5) becomes small due the small mean spectrum in the corresponding **W** column or due to the small mean image in the **W** row. A simple derivation shows that $\sigma^2 \to \frac{mn}{c}$ in the "saturated" limit and the noise variance can be successfully rescaled over data matrix **D** to match this limit. In the "diluted" limit, the noise variance is constrained as $\sigma^2 \le mnc$, which causes the failure of the noise equalization.

Appendix 2 presents simulations of STEM XEDS datasets for a simplified one-component object where the signal strength was varied in the large range. The extracted mean noise variance fits nicely the theoretical consideration of Appendix 1.

## Appendix 2: Simulation of the effect of sparsity on weighting XEDS spectrum-images for a one-component object

A simple object consisting of two phases - pure Si and $SiO_2$ with the smooth variation of composition between them—was employed in the simulations. The XEDS spectra were generated in the same way as described in "Details of simulation" section. The synthetic spectrum-images consisted of $100 \times 100$ pixels and 300 energy channels covering the energy range 0–3 keV.

The simulations were carried out with a systematic variation of the total number of counts in the spectrum-image or in other words, with a varying dose. As in "Multi-component object for spectrum-imaging" section, each dataset was generated in two variants—with and without adding a Poissonian noise. The following datasets were created: $\frac{1}{16}, \frac{1}{8}, \frac{1}{4}, \frac{1}{2}, 1, 2, 4, 8, 16, 32, 64$, where the number denoted the dose. Dose 1 corresponds to the conditions for the simulations in "Details of simulation" section, i.e. a typical dose encountered in STEM XEDS spectrum-imaging nowadays.

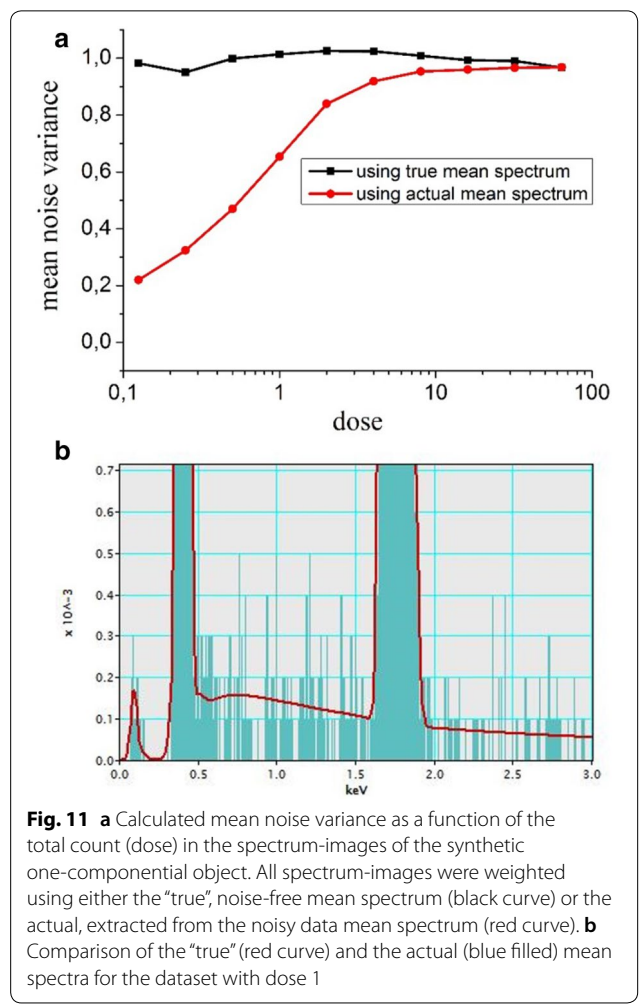

**Fig. 11 a** Calculated mean noise variance as a function of the total count (dose) in the spectrum-images of the synthetic one-componental object. All spectrum-images were weighted using either the "true", noise-free mean spectrum (black curve) or the actual, extracted from the noisy data mean spectrum (red curve). **b** Comparison of the "true" (red curve) and the actual (blue filled) mean spectra for the dataset with dose 1

Then, the synthetic spectrum-images were weighted by normalizing it to the mean spectrum (spectrum-only weighting). The latter was extracted either from the noisy dataset or from the corresponding noise-free dataset. Finally, the total variance of noise was evaluated by subtracting the noisy datasets from the noise-free ones.

Figure 11a demonstrates that the weighting procedure normalizes noise to 1 only if the "true" mean spectrum is utilized. Using the actual mean spectrum causes the weighting artifacts at low doses. As seen from Fig. 11b, the mean spectrum varies very discretely there and this uncertainly propagates to the weighted datasets causing uneven distribution of the noise variance over the different energy channels.

Unfortunately a "true" mean spectrum is not available in real experiments. Using instead a mean spectrum evaluated from a noisy dataset always imposes a risk of failure of the noise equalization as described in "Appendix 1".

### Appendix 3: Scatter plots of the principal component decomposition for the experimental and synthetic datasets

Figure 12 shows the results of the weighted principal component decomposition for the experimental and synthetic datasets expressed in the scatter plots of component. In contrast to real-space images or maps, scatter plots visualize the data distribution in a kind of spectral or factor space. The spatial positions of data points plays no role in scatter plots. Instead, the appearance of data in such plots is determined solely by the spectral proximity of pixels constituting a spectrum-image. The eigenspectra of the latent factors governing the data variation can be sometimes identified as special knots in this spectral space.

A benefit of noise-free synthetic data (Fig. 11c, f, i, l) is in clear visibility of the data variation among given latent factors. The variation trend might represent a more or less straight line connecting the two factor knots through the shortest way or might follow some bent curve. Typically, the latter behavior indicates non-linearity in the spectra formation [25]. The introduction of noise (Fig. 11b, e, h, k) results in smearing out the true scatter plots such that the only dominant variation trends are visible.

The scatter plots of the experimental dataset (Fig. 11a, d, g, j) are in a good agreement with those for the noisy synthetic dataset (Fig. 11b, e, h, k). The slight discrepancies in the positions of the latent factors and the shape of variations can be attributed to the simplified assumptions in the spectra modeling and to the limited accuracy in reproducing the actual structure of the CMOS device. Note that the scatter plots represent the 2D cross sections in a spectral space with 1200 dimensions, thus even small deviations in the orientation of the given cross section might influence noticeably the appearance of the data simplex.

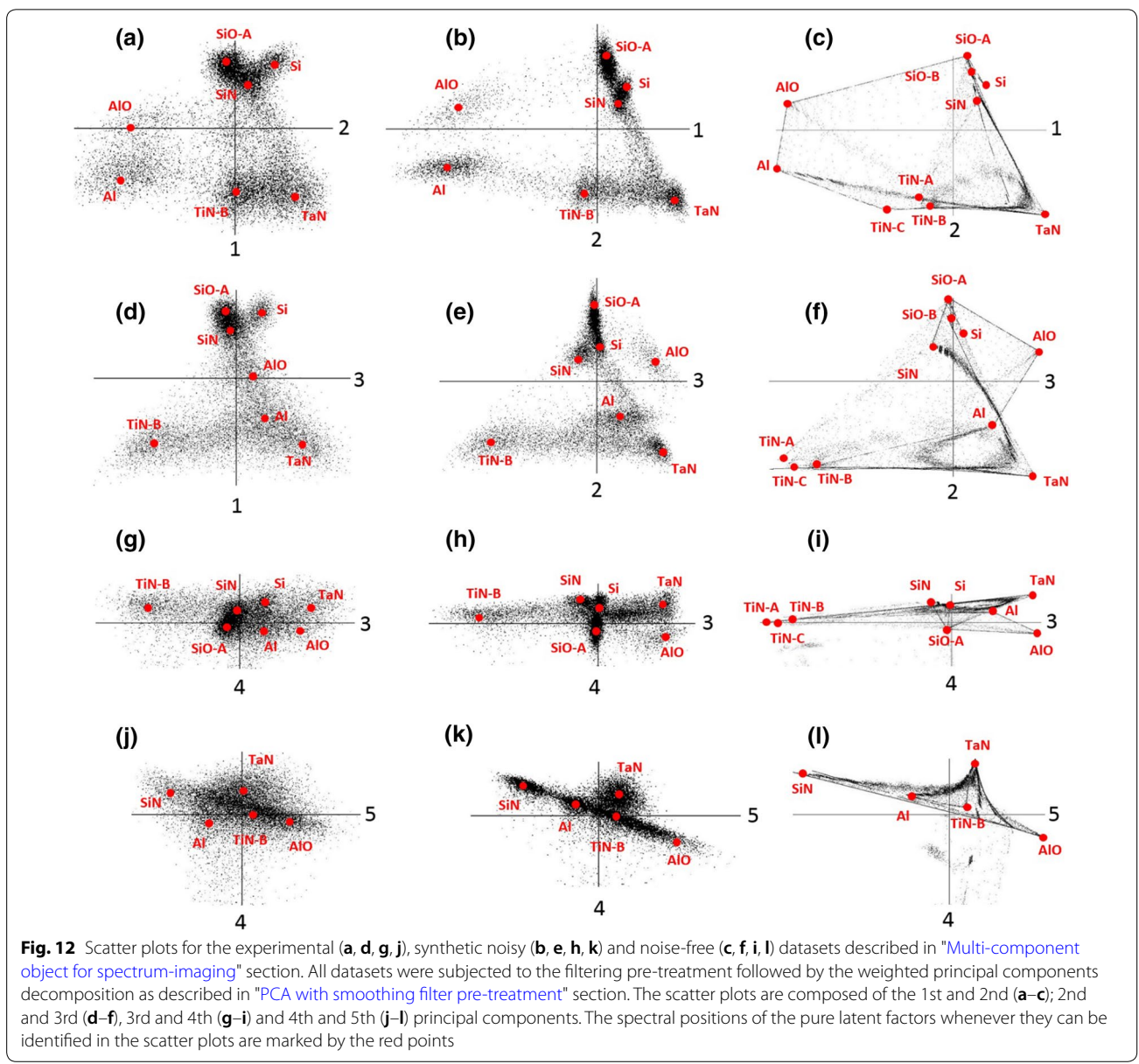

**Fig. 12** Scatter plots for the experimental (**a**, **d**, **g**, **j**), synthetic noisy (**b**, **e**, **h**, **k**) and noise-free (**c**, **f**, **i**, **l**) datasets described in "Multi-component object for spectrum-imaging" section. All datasets were subjected to the filtering pre-treatment followed by the weighted principal components decomposition as described in "PCA with smoothing filter pre-treatment" section. The scatter plots are composed of the 1st and 2nd (**a–c**); 2nd and 3rd (**d–f**), 3rd and 4th (**g–i**) and 4th and 5th (**j–l**) principal components. The spectral positions of the pure latent factors whenever they can be identified in the scatter plots are marked by the red points

The proximity between the experiment and simulations is however deteriorated with increasing the component index. This happens because the the variance of the high-index components is reduced and the accuracy of the retrieved component loadings is reduced as well.

**Appendix 4: Examples of application of the anisotropy method for truncating principal components in STEM XEDS spectrum-images**

The STEM XEDS data were collected under the conditions described in "Experimental details" section with similar beam currents and acquisition times. To reduce the data sparsity the spectrum-images were fist subjected to $2 \times 2$ binning and then to Gaussian kernel filtering with the standard deviation $\sigma = 1$ pixel.

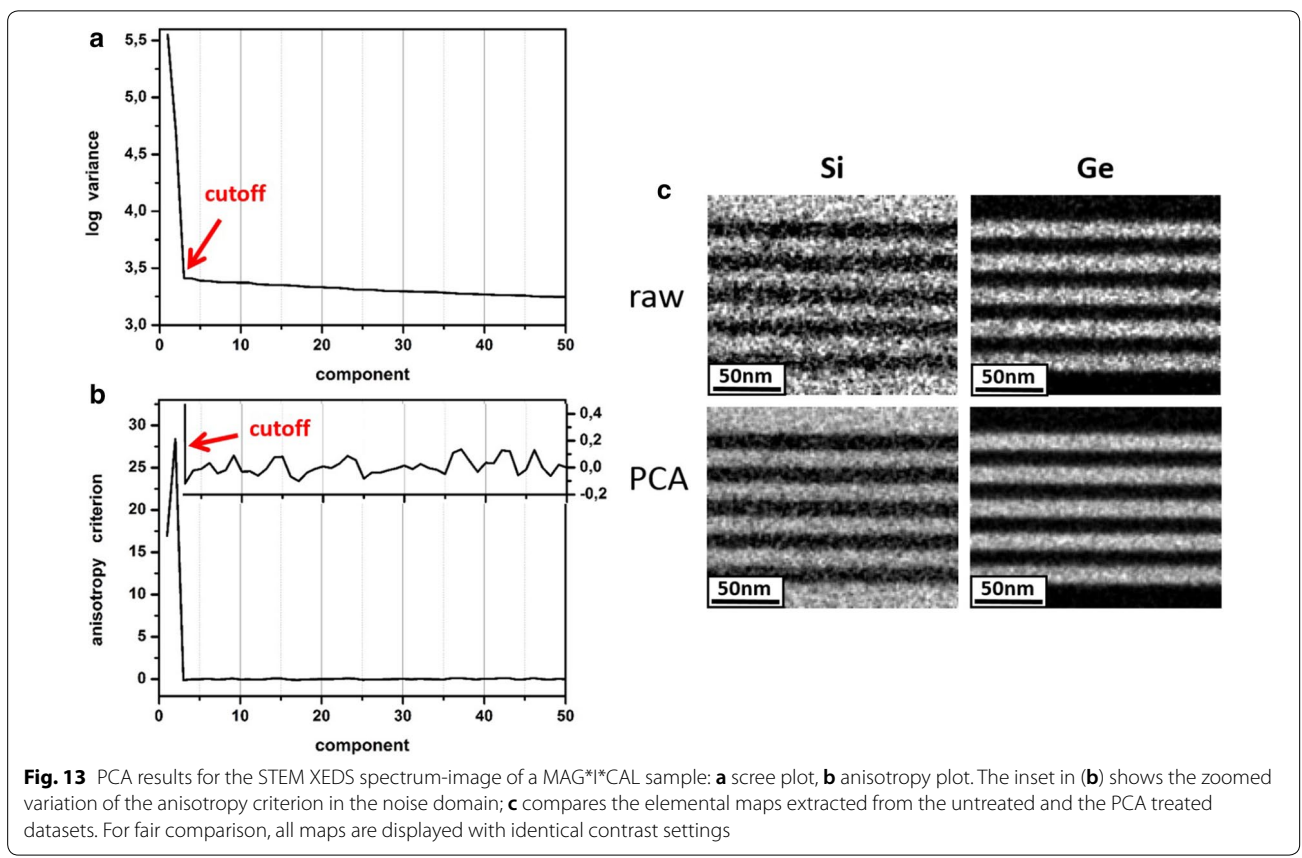

**Fig. 13** PCA results for the STEM XEDS spectrum-image of a MAG*I*CAL sample: **a** scree plot, **b** anisotropy plot. The inset in (**b**) shows the zoomed variation of the anisotropy criterion in the noise domain; **c** compares the elemental maps extracted from the untreated and the PCA treated datasets. For fair comparison, all maps are displayed with identical contrast settings

Figure 13 presents the evaluation of well-known MAG*I*CAL samples dedicated for a magnification calibration in the TEM. The active structure of MAG*I*CAL is composed of a nanoscaled stack of Si and Si-15%Ge layers that were characterized by STEM XEDS followed by the application of PCA. As a result, the elemental maps of Si and Ge were drastically denoised and improved as evident from Fig. 13c.

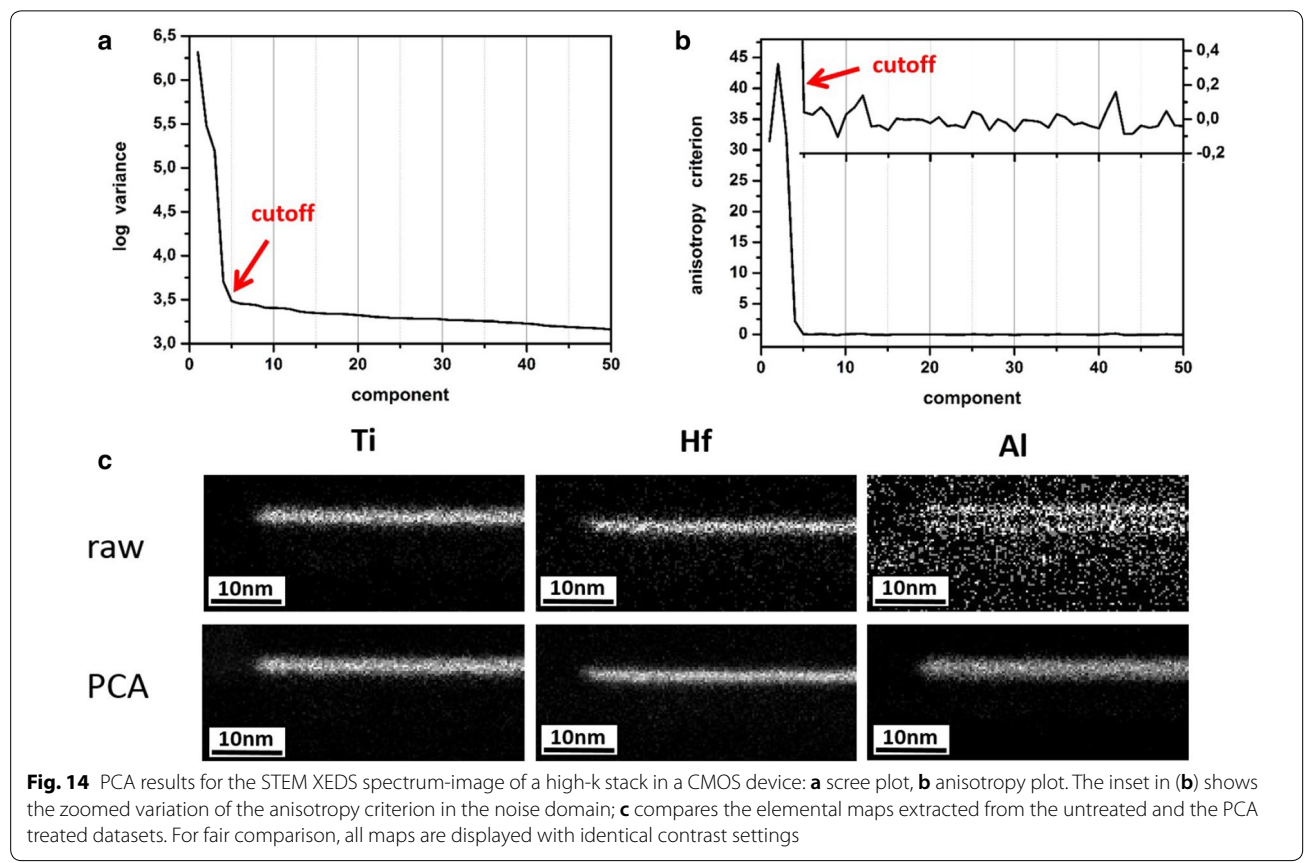

**Fig. 14** PCA results for the STEM XEDS spectrum-image of a high-k stack in a CMOS device: **a** scree plot, **b** anisotropy plot. The inset in (**b**) shows the zoomed variation of the anisotropy criterion in the noise domain; **c** compares the elemental maps extracted from the untreated and the PCA treated datasets. For fair comparison, all maps are displayed with identical contrast settings

**Table 4 The number of components to truncate according the different truncation methods: the evaluation of a scree plot with visual localisation of the inflection point ("Scree plot method for truncation of principal components" section), the approach of Gavish and Donoho ("Analytical model-based methods for truncating principal components" section) and the anisotropy method ("Anisotropy method for truncation of principal components" section) with using the projected histograms and the anisotropy threshold of 0.5**

| Dataset | Scree plot | Gavish and Donoho | Anisotropy |
|---|---|---|---|
| MAG*I*CAL | 2 | 2 | 2 |
| High-K stack | 4–5 | 49 | 4 |

Figure 14 shows the results of the PCA treatment for the so-called "high-*k* stack" in modern semiconductor CMOS devices. This stack serves to maximize the dielectric constant *k* and therefore to increase the gate capacitance. In the considered case, the high-*k* stack consisted of a $HfO_2$ layer capped by TiN where both layers were doped with Al. The PCA treatment changes only slightly the appearance of the strong Ti and Hf signals but improves dramatically the visibility of the weak Al signal in Fig. 14c.

Table 4 compares the performance of the different methods for truncating principal components. The scree plot and anisotropy methods suggest a similar number of meaningful components although the anisotropy method is a bit more robust and precise. The analytical method of Gavish and Donoho suggests the same cut-off for the MAG*I*CAL sample while overestimates heavily the number of meaningful components for the high-k stack sample.

### Publisher's Note